\documentstyle[12pt]{article}

\begin{document}


\hsize=6.15in
\vsize=8.2in
\hoffset=-0.42in
\voffset=-0.3435in

\normalbaselineskip=24pt\normalbaselines

{\bf  Published in:  ``Mathematica Applicanda'' 40: 27-37 (2012) }

\vspace{3.5cm}

\begin{center}
{\large \bf What can a mathematician do in neuroscience?}
\end{center}

\vspace{0.15cm}

\begin{center}
{Jan Karbowski$^{a,b,*}$}
\end{center}

\vspace{0.05cm}

\begin{center}
{\it $^{a}$ Institute of Biocybernetics and Biomedical Engineering, \\
Polish Academy of Sciences, 02-109 Warsaw, Poland \\
$^{b}$ Institute of Applied Mathematics and Mechanics, \\
University of Warsaw, 02-097 Warsaw, Poland }
\end{center}


\vspace{0.1cm}

\begin{abstract}
Mammalian brain is one of the most complex objects in the known universe,
as it governs every aspect of animal's and human behavior. It is fair to say 
that we have a very limited knowledge of how the brain operates and functions. 
Computational Neuroscience is a scientific discipline that attempts to 
understand and describe the brain in terms of mathematical modeling. 
This user-friendly review tries to introduce this relatively new field
to mathematicians and physicists by showing examples of recent trends.
It also discusses briefly future prospects for constructing an integrated 
theory of brain function.

\end{abstract}




\noindent {\bf Keywords}: Computational Neuroscience; Brain, Modeling, 
Neurons.

\vspace{0.1cm}

\noindent $^{*}$ Corresponding author at: jkarbowski@mimuw.edu.pl; 
jkarbowski@ibib.waw.pl.

\vspace{0.3cm}

\newpage

%

\section{Introduction}

The purpose of this review article is to present a relatively new field
of Computational Neuroscience (or Theoretical Neurobiology) to mathematicians
or physicists, who would like to do a non-traditional research in theoretical
biology but do not know how and where to start. The primary audience for
this review are graduate students who have their MSc in mathematics, theoretical
physics, or computer science, and are ambitious enough to think about their 
PhD (doctorate) in Computational Neuroscience.

I still remember myself in the mid 1990s when I was finishing my PhD in
theoretical physics (condensed matter) and thinking that I would not spend
the rest of my life doing this type of research. At that time it became clear
to me that physics was an ``old science'' and all exciting theoretical
problems that could be solved and later verified experimentally had been already
solved. (Of course there was string theory but that appeared to me as an extremely
abstract fantasy with its 11 or so dimensions and with no chance for any sort
of verification in my lifetime). In my search, I turned to biology, which
at about that time was depicted in a popular press as the science that would
dominate the 21 century. Judging by the percent of biological papers published
in every issue of highly prestigious Nature and Science magazines, this
prediction has turned out to be correct. In particular, neurobiology seemed
very interesting to me, because it dealt with the brain, which is the organ
generating all of our behavior, as well as higher cognitive
states (e.g. the ability to solve math equations). I made my transition from
theoretical physics to theoretical neurobiology when I was a postdoc at
Boston University. That transition was relatively quick and painless, and
I have never regretted my decision. My example shows that it is possible to
switch successfully scientific fields, if somebody is highly motivated and
hard working. Thus, if a young reader of this article possesses these two
traits, such a transition can be possible as well.

\section{General overview of Computational Neuroscience}

\subsection{Grand challenges of Computational Neuroscience}

One of the major goals of contemporary Neuroscience is to understand human
behavior and action through understanding certain physical processes in
the brain \cite{arbib}. The challenge here is to provide quantitative description 
or ``theory'' that would make far-reaching testable predictions, much the same
way as it has happened in physical sciences with the understanding of
non-living matter. There is no need to elaborate that such putative
understanding would be beneficial for society at large, and might have
technological implications (e.g. for constructing ``intelligent''
computational devices). However, because of the brain structural complexity
(about $10^{10}$ neurons and $10^{15}$ synapses in the human brain),
this task is extremely difficult. One might note that this fact in itself
should not be a big obstacle because even 1 mole of non-living matter
contains about $10^{23}$ atoms, and somehow statistical physics deals efficiently
with that. However, there are several fundamental differences between
non-living and living matter, in particular the brain.

First, the brain, which can be viewed as an electro-chemical system of
ionic mixture, is far from thermodynamic equilibrium with the environment and 
also within different brain components.
This inequilibrium is maintained by the brain itself through various
self-regulatory biochemical and sensory feedback loops, which use energy
from the environment (animals have to eat). Nothing like that appears
in non-living matter composed of a huge number of atoms, and mostly successfully
described by equilibrium statistical physics (there are also exceptions, e.g.
whether phenomena that are often described by non-equilibrium thermodynamics).

Second, neurons interact in a non-linear manner and the degree of this
interaction changes over time (neuromodulation). Moreover, neurons and 
synapses undergo structural changes over vastly different time scales (from 
minutes to years). This process is known as brain plasticity and it is to 
a large extent environmentally (externally) driven, and therefore it has 
a strong non-deterministic component. This environmental stochasticity introduces
non-stationarity into brain dynamics, which is difficult to capture
theoretically. In contrast, non-living physical systems described by 
equilibrium statistical physics are composed of elements that interact 
in a predictable and on average static ways, which are relatively easy 
to formulate mathematically.

Third, neurons process information, i.e. they store information and
recall it when necessary. A similar function can be also performed
by some non-living man-made devices, e.g. computers, but only rather passively. 
The point is that brains do it naturally and adaptively, with a high degree
of effectiveness, which is a result of a long process of evolution and self-organization
(see e.g. \cite{kaufmann} for the latter topics). A related issue is that brains
(at least human brain) use information in a specific way to create abstract
representations of the outside world. This is implemented by the so-called 
``higher cognitive states'' or consciousness, which form the basis of our daily 
existence and experience. Unfortunately, these cognitive states are rather
elusive and therefore they have been much more investigated using the methods
of psychology rather than those of exact sciences.

Given all that, one could speculate that the future integrated ``grand theory'' 
of brain function (provided such a theory is possible at all) should contain 
elements of stochastic dynamical systems \cite{strogatz,gardiner} combined in 
some innovative way with non-equilibrium statistical physics \cite{nicolis}, and 
information theory \cite{hofkirchner} possibly with some elements of game theory 
(psychologically motivated). All these three or four disciplines already exist 
and have solid theoretical foundations. Nevertheless, they are still actively 
developed and have many open questions, especially in the context of neuroscience. 
Thus, I believe that we have enough theoretical tools at hand for describing brain 
functional mysteries, and there is no need to invent ``new kinds'' of mathematics 
or physics for that particular goal (see however \cite{deschutter} and \cite{wolfram} 
for the opposite points of view on the brain and general complex biological 
systems, respectively). It seems that the main challenge in constructing the grand
theory of the brain is in appropriately adopting and integrating already existing 
separate concepts from these disciplines into a coherent theoretical picture that 
would be useful for brain description, rather than to invent completely 
new and unchecked concepts.

\subsection{Historical remarks}

Up to recent years Neuroscience has been predominantly an experimental
science, in which scientists have been accumulating painstakingly, mostly
separate, experimental facts.
In this field there was no room for any sort of theory or mathematical
model, partly because of incomplete or evolving knowledge and partly because
of neuroscientists' reluctance to mathematics. Consequently, the majority
of neurobiological models had qualitative, verbal character.
The first example that theory can be useful in neuroscience came with
the formulation of the so-called Hodgkin-Huxley model \cite{hodgkin}, which
``mechanistically'', i.e. physically, explains the generation and propagation
of action potential in squid axon in terms of Na$^{+}$ and K$^{+}$ ions flow 
through neuron's membrane. That mathematical model agreed perfectly with the 
data and subsequently has become a classic. Hodgkin and Huxley were later 
awarded the Nobel Prize in Medicine or Physiology (1963) for their combined 
experimental, computational and explanatory efforts of neural spiking biophysics. 
However, despite this early success, the mathematical modeling in general and the 
Hodgkin-Huxley model in particular, were for a long time an exception rather than 
a rule in the neuroscience research. The exceptions in the 1960s and 1970s were 
in part due to people like Stephen Grossberg, Walter Freeman, and slightly 
later Terry Sejnowski (to name just a few), who made some lasting contributions
to computational neuroscience, but nevertheless, did not inspire big
crowds of computational scientists. Such an inspiration wave among
physicists came later with the so-called Hopfield models in the early 1980s
\cite{hopfield}. That huge initial wave subsided significantly later
because the Hopfield's models and their extensions were not too realistic.
It seems that the main value of these simple models is in the fact that they
helped to realize that mathematics, physics, and computer science with their
quantitative and rigorous methodologies, can offer a lot in understanding
the brain. As a result of this thinking a new field called Computational
Neuroscience has emerged, which is still in progress and which provides
a link between theoretical and experimental work in neuroscience from
a physical perspective \cite{koch,dayan}. It seems that at present Neuroscience 
has matured enough so that it is possible, and even necessary, to try to find
mechanistic explanation of brain dynamics and function.
The current models used in Computational Neuroscience are far more
complex and realistic than they were in the 1980s and 1990s.
It should be also mentioned that this new interdisciplinary field
is practiced only to a limited extent in Poland by a handful of people.
Nevertheless, I do hope that a popularity of this exciting field will grow 
over time in Poland as well.

\section{Examples of research topics in Computational Neuroscience}

In this section I present a short description of selected topics
in Computational Neuroscience that might be of a particular interest
to the mathematical and physical communities. The choice of these
topics reflects author's interest, and for that reason it should
not be viewed as the whole field of Computational Neuroscience.

\subsection{Dynamics of a single neuron and networks of neurons}

The most popular area of research in Computational Neuroscience is
neural dynamics. Neurophysiological studies show that the brain exhibits 
different activity patterns, from regular oscillations, often with synchronous 
activities across different brain regions, to highly irregular or chaotic 
behavior. In this area the typical questions of interest are:
(i) What are the mechanisms of oscillations in neural systems?
(ii) What conditions must be satisfied to obtain synchronization
in neural activities? (iii) Is chaotic neural activity relevant functionally?
(iv) What is the mechanism of generating the so-called bursts of action
potentials? These and similar questions are investigated both on a single 
neuron level and on a network level. In general, the single neuron study 
involves modeling realistic neurons with complicated voltage dynamics 
due to primarily sodium and potassium channels located on neuron's
membrane. The basic equations of this approach, called
Hodgkin-Huxley (HH) equations are of the form:

\begin{equation}
C\frac{dV}{dt}= -g_{L}(V-V_{L}) - g_{Na}(V-V_{Na}) - g_{K}(V-V_{K}) + I_{syn},
\end{equation}\\
where $C$ is the membrane electric capacity, $V$ is the membrane voltage, 
$I_{syn}$ is the synaptic input current, 
$g_{Na}$, $g_{K}$, and $g_{L}$ are sodium, potassium and the so-called leak 
conductances through the membrane. These ions have their specific equilibrium
voltages (Nerst potentials) for balance of their concentration gradients 
with electrostatic forces. These voltages are denoted as $V_{Na}$, $V_{K}$, 
and $V_{L}$, respectively. The Na and K conductances are dynamical parameters 
in this model, and are given by $g_{Na}= \overline{g}_{Na}m^{3}h$, and
$g_{K}= \overline{g}_{K}n^{4}$, where $\overline{g}_{Na}$ and 
$\overline{g}_{K}$ are maximal conductances (all positive). The variables 
$m$, $h$, and $n$ are the so-called gating variables, and each of them 
is described by a similar differential equation of the type: 

\begin{equation}
\frac{dm}{dt}= \alpha_{m}(1-m) - \beta_{m}m,
\end{equation}\\
where $\alpha_{m}$ and $\beta_{m}$ depend on voltage $V$ in a non-linear
manner. The gating variables describe complicated ion channels kinetics 
such as channel opening, closing, and inactivation through voltage dependence 
of the parameters $\alpha$ and $\beta$ (for details see \cite{koch,dayan}).

Let us try to provide some physical picture behind Eqs. (1) and (2).
When neuron does not get a synaptic input ($I_{syn}= 0$), its voltage $V$
is at a resting potential, which is close to $V_{L}$ (about $-65$ mV).
This is a consequence of the fact that for very negative voltages,
Na$^{+}$ and K$^{+}$ channels are practically closed and do not conduct
ions. Mathematically this means that the gating variables $m$ and $n$,
characterizing channels openings, are approximately to zero (hence $g_{Na}$
and $g_{K} \rightarrow 0$). However, when synaptic input $I_{syn} > 0$,
then voltage $V$ increases due to sodium channels opening and Na$^{+}$
influx to neuron's interior. Mathematically speaking, $V$ grows because
$g_{Na}$ increases and $V-V_{Na} < 0$ ($V_{Na}\approx 50$ mV). 
If $I_{syn}$ is sufficiently strong, then $V$ can reach a threshold
($V_{th}\approx -44$ mV) for generation of an action potential (action
potentials are abrupt changes ``spikes'' in $V$, and are the means to
communicate signals between neurons). After crossing the threshold, the
voltage increases further sharply, and this is because a positive feedback
loop between $g_{Na}$ and $V$ (both drive each other through the gating
variable $m$). With some delay the potassium channels start to open, i.e.
the gating variable $n$ starts to grow. However, since $V-V_{K} > 0$
($V_{K}\approx -90$ mV), the potassium current is negative (K$^{+}$
ions escape from the neuron's interior) and counteracts the positive
Na$^{+}$ current. Therefore, the activation of potassium channels results 
in slowing down the rise of $V$, and ultimately its decay after reaching
$\sim 20-30$ mV to negative values below $V_{L}$. From there the voltage
slowly relaxes to its resting value $V_{L}$. The whole process
of the action potential lasts about $2-3$ msec.

Note that the Hodgkin-Huxley HH model involves four differential equations
(one for $V$, and 3 equations for the gating variables). The modern single 
neuron research takes this model as a base and extends it by including 
a whole range of new (recently discovered) channels, such as different types 
of calcium channels, calcium activated potassium channel, etc, to study neural 
dynamics. Such models are more realistic but at the same time more complex (and 
require more computing time in simulations). The interesting thing is that
sometimes the presence of one channel type can have a dramatic impact on single 
neuron dynamics \cite{achard,koch2}. That is, this dynamics can vary
from highly ordered to extremely irregular.

On the other hand, studies on a network level involve simplified models of
neurons. The most popular of these is the so-called integrate-and-fire
(IF) model, of the form

\begin{equation}
\tau\frac{dV}{dt}= f(V) + I_{syn},
\end{equation}\\
where $\tau$ is the membrane time constant, and the function $f(V)= -(V-V_{r})$, 
with $V_{r}$ denoting the resting membrane potential (i.e. when there is
no synaptic input). In this model an action potential is generated when
voltage $V$ reaches a certain threshold, after which $V$ is immediately reset
by hand to a value below $V_{r}$. Then $V$ starts to relax back to value $V_{r}$.
This resetting procedure introduces a discontinuous jump in $V$, which is to
mimic the decay phase of $V$ during an action potential observed in real neurons.
Additionally, the resetting jump in $V$ leads to a non-linearity in the model, 
which otherwise would be perfectly linear. The advantage in 
using this simple model over the HH type model (Eqs. 1 and 2) is 
that a simulation time on a computer is much shorter because less equations 
have to be solved. This allows simulation of a huge number of connected IF 
neurons. Another benefit of using IF model is that, unlike HH model, it can 
be solved analytically, which provides in some cases a huge intuitive 
advantage. On the other hand, the drawback of this model is that ``an interesting''
neural dynamics is generated only by an interesting synaptic input, unlike
in HH base models (which could be highly nonlinear even with constant $I_{syn}$).

Extensions of IF model have been proposed that involve different forms
of the function $f(V)$. For example, the so-called quadratic integrate-and-fire
or its modification known as an Izhikevich model has $f(V)= aV^{2}+bV+c$,
where $a, b$, and $c$ are some numerical coefficients 
\cite{brunel,izhikevich}. Another example was proposed by the author 
\cite{karbowski2000} that can be called an absolute integrate-and-fire, which 
is piece-wise linear with $f(V)= |V-V_{r}|$. These extended models have features 
that make them slightly more realistic than the traditional IF model, and hence
should describe the dynamics of real neural networks a little more faithfully.

\subsection{Models of learning and memory in the brain }

One of the most important aspects of the brain is that it can learn and
remember different events in the real world. These dynamic processes take
place in the synapses, i.e. in the tiny volumes connecting two neurons (the
term $I_{syn}$ in Eqs. 1 and 3). Despite a huge experimental progress in the 
last 30 years, the detailed mechanisms of learning and memory are still 
poorly known. What we do know, however, is that synaptic structure and 
conductance are not static but change over time with different time constants 
ranging from 0.1 sec to $\sim 50-70$ years (human lifetime). That process is 
known as synaptic plasticity, and it can be activated when two neurons 
connected by a given synapse are simultaneously (or almost simultaneously) 
active. It is suspected that memories are encoded in these structural
synaptic changes.

There exist a traditional model of memory based on the Hopfield
model \cite{hopfield}. In this model the network learns different patterns
by a training, and there is a close correspondence of memory states to
the basins of attractors known from the dynamical systems. The problem with
this and similar models is that they are quite abstract, i.e. memories
can last forever, and moreover they have rather low capacity for the number 
of memory patterns they can store. In recent years, there have appeared other 
models of learning and memory \cite{amit,fusi,leibold} that are based on an 
experimental finding that synapses can exist in many discrete states, not 
in a continuum of states \cite{montgomery}. In these models 
\cite{amit,fusi,leibold}, which usually have large memory capacities, plasticity 
is associated with transitions between these discrete states, and memories can 
naturally fade away, as it happens in the real world. As was stated above,
there is a stochastic component to these transitions, which involves
considering probabilities of synaptic states. The dynamics of synaptic probabilities 
are described by differential equations of the Master equation type known 
from non-equilibrium statistical physics \cite{gardiner}. Anther formulation of 
synaptic plasticity through the Fokker-Planck equation is also possible. 
This example shows that stochastic effects are present in the brain and can have 
some functional role (in this case in learning and memory).

\subsection{Models of brain metabolism and visualization of brain function }

Brain is an expensive organ in terms of metabolic energy it uses 
\cite{aiello,attwell}. Moreover, as brain increases in size on an evolutionary
scale its metabolic consumption grows slightly faster than metabolic needs
of the whole body \cite{karbowski2007,karbowski2011}. This suggests that
metabolism is an important part of brain function. The majority of energy
used by neurons goes to maintaining concentration gradients of Na$^{+}$
and K$^{+}$ ions across neural membranes. These gradients are necessary
for keeping the brain in the out of equilibrium state, which is the
pre-condition for the ability to generate action potentials, which are
necessary for efficient neural communication. One can relate the amount of 
metabolic energy used to the underlying neurophysiological processes such as 
the firing frequency of action potentials and activities of synapses. This
involves solution of a system of differential equations for a balance in
ionic flow of Na$^{+}$ and K$^{+}$ \cite{karbowski2009}. This relationship 
can have a practical aspect, because by measuring metabolic activity of 
certain brain regions one could say something more definitive about 
physiological state of neurons there. This is important, since the only way 
to visualize regional brain function in humans is through techniques such 
as PET (positron emission tomography) and fMRI (functional magnetic resonance 
imaging). Both of them are based primarily on brain metabolic activity,
and therefore it is good to have a model that maps brain metabolism into
an underlying electrical (neural) and chemical (synaptic) signaling.
In my opinion, this branch of neuroscience will develop fast in the coming
years, as is it visible in recent conferences and workshops agendas (e.g. Computational
Neuroscience Meeting in Stockholm CNS 2011, or INCF Congress in Munich 2012).

\subsection{Models of brain connectivity patterns }

The mammalian brain is organized hierarchically. On the most basic level, 
there are neurons that are connected by synapses. Typically, there are
two major classes of neurons, excitatory and inhibitory, determined by
the type of synapses (neurotransmitters) they make with other neurons.
It is commonly believed that neurons are organized into the so-called columns 
(about $10^{3}-10^{4}$ neurons). Neurons belonging to a single column behave 
dynamically similar in response to a specific stimulation, and different columns 
are activated differentially by the same stimulus. Columns are organized into 
macro-columns and these, in turn, form functional ``areas''. Human brain contains 
about hundred areas, each one is thought to process a different type of information.
For example, visual areas process visual input coming to the brain through
the eyes. Motor areas guide movement of hands and legs, whereas frontal
areas (located in front of the brain) are associated with higher cognitive
functions.

If the brain is to perform its functions coherently, different areas have
to communicate efficiently, i.e. on time, so that information is globally
integrated \cite{laughlin,karbowski2003}. This means that a certain level of 
connectivity between neurons and areas should be maintained. Too low 
connectivity would imply too large separation between areas, which is not 
good for integration of information.
On the other hand, too high connectivity would unnecessarily merge distinct
areas causing their functional disintegration. Studies in recent years
have shown that brain has a ``small world'' architecture \cite{bassett},
which is a name given to organization in which there are many local
connections and only few long-range connections \cite{watts}.

A common theoretical tool for investigating these topics is graph theory,
which provides quantitative means for characterizing anatomical and functional
separation and integration in a network \cite{bullmore2009}. Research in this
field is centered around questions of how to relate brain functioning to
parameters characterizing brain connectivity. For example, there are
studies claiming that certain brain abnormalities and disorders, like
schizophrenia and autism, may be a result of an altered connectivity between 
brain regions \cite{bullmore2012}.

\section{Concluding remarks}

It is important to stress a difference between understanding the main ideas
of the field of Computational Neuroscience and doing an ambitious research
on theories of brain function. For the former, one does not need overly complicated
or sophisticated mathematics. In fact, it seems that the most common theoretical
tool across the whole theoretical neuroscience is the ability to solve and
analyze systems of non-linear differential equations on a computer (simulations).
By mastering this relatively simple technique (there are even open source free
softwares that do it for you), one acquires in principle a sufficient technical
background to comprehend the basics of theoretical neuroscience. For a general 
overview of mathematics used in neuroscience see the book by Ermentrout and
Terman \cite{ermentrout}.

On the other hand, constructing influential theories of brain functioning
requires in my opinion a mixture of different skills (as it was signalled
in Section 2). It is not enough to just know how to solve differential equations
on a computer, one has also to know how to analyze such equations analytically,
at least approximately and qualitatively, to gain an intuition about what 
they really describe. The more analytical techniques one knows the better. 
However, even this can be insufficient without a solid knowledge of physics and 
traditional neurobiology. Physics teaches how to describe in mathematical terms 
the non-living matter, and borrowing certain concepts from this discipline may be 
highly beneficial. In particular, the ideas of non-equilibrium thermodynamics 
and electromagnetism applied in a new context to the brain may yield new insights 
of how the brain works. Knowledge of neurobiology on some decent level should also 
help. Neurobiology (or biology in general) teaches researchers from the so-called 
``exact sciences'' which processes are possible and which are not, and additionally 
that biological phenomena are generally governed by many factors (or parameters), 
often conflicting in outcome.

The current Computational Neuroscience is very fragmented. Virtually almost all
theoretical papers in this field focus on a specific neurobiological phenomenon
or on a narrow class of phenomena (no exception are the examples presented in
Sec. 3). This narrowness of the scope probably reflects the degree of difficulty
in constructing theories that would have a broader unifying impact. This
difficulty is a consequence of the fact, already mentioned in Sec. 2, that 
``grand theories'' require a comprehensive approach that would have a chance
to integrate concepts from different disciplines into a coherent framework
with testable (quantitative) predictions. That is probably too much for the
present-day Computational Neuroscience. For example, it is currently not clear
at all how to pass from molecular (neurotransmitters, channels) and cellular
(neuron) levels of description to the description of macroscopic cognitive
states. I do not think we can even properly address this question at present.
The point is that we are unsure about which microscopic details to include
and which to abandon in the theoretical description. Similarly, on the
macroscopic level, it is often unclear how to precisely define distinct
cognitive states, and which parameters should characterize them.

The research topics discussed in Sec. 3 and almost all other (see \cite{arbib}) 
are satisfactory described in terms of classical physics. However, since neurons 
and synapses are so small (micrometers), one can wonder if quantum mechanics 
(describing the world of atoms and molecules) is relevant for the brain description 
(I mean here functional or dynamical description not just a mere structural description 
of various microscopic parts of channels and synapses). About two decades ago 
a prominent mathematician Roger Penrose published a book \cite{penrose} in which he 
claimed that quantum effects might underlie neural activities that lead to 
consciousness. The book was totally criticized by almost every neuroscientist 
for the author's lack of basic neurobiological knowledge, but nevertheless caused 
the people to ask questions about the relationship between quantum theory and brain 
function \cite{tegmark,hameroff,koch2006}. At present, it is rather commonly believed 
that quantum coherence effects are too short ($10^{-10}-10^{-15}$ sec) to be important
for brain dynamics ($10^{-3}$ sec), and thus they cannot control brain function. 
The very short quantum coherence times are caused by frequent ionic (Na$^{+}$, K$^{+}$, 
Cl$^{-}$, Ca$^{++}$) collisions with each other and water molecules, which is additionally 
amplified by a relatively high brain temperature of 300 K \cite{tegmark,koch2006}.

\vspace{1.3cm}

\noindent{\bf Acknowledgments}

The work was supported by the grant from the Polish Ministry of Science
and Education (NN 518 409238), and by the Marie Curie Actions EU grant 
FP7-PEOPLE-2007-IRG-210538.

\newpage

\vspace{1.5cm}



\end{document}